\documentclass[9pt,conference]{IEEEtran}
\IEEEoverridecommandlockouts

\usepackage{cite}
\usepackage{amsmath,amssymb,amsfonts}
\usepackage{algorithmic}
\usepackage{graphicx}
\usepackage{textcomp}
\usepackage{xcolor}
\usepackage{hyperref}
\hypersetup{
hidelinks,
colorlinks=true,
linkcolor=black,
citecolor=black,
urlcolor = black
}
\usepackage{booktabs,multirow,makecell}
\def\BibTeX{{\rm B\kern-.05em{\sc i\kern-.025em b}\kern-.08em
    T\kern-.1667em\lower.7ex\hbox{E}\kern-.125emX}}
\begin{document}

\title{Enabling Auditory Large Language Models for\\ Automatic Speech Quality Evaluation}

\author{
\IEEEauthorblockN{Siyin Wang$^{1}$, Wenyi Yu$^{1}$, Yudong Yang$^{1}$, Changli Tang$^{1}$, Yixuan Li$^{1}$, Jimin Zhuang$^{1}$ \\  Xianzhao Chen$^{2}$, Xiaohai Tian$^{2}$, Jun Zhang$^{2}$, Guangzhi Sun$^{3}$, Lu Lu$^{2}$, Yuxuan Wang$^{2}$, Chao Zhang$^{1 *}$\thanks{$^{*}$Corresponding author.}}
\IEEEauthorblockA{
\textit{$^{1}$Tsinghua University},
\textit{$^{2}$ByteDance},
\textit{$^{3}$University of Cambridge} 
}
\textit{wangsiyi23@mails.tsinghua.edu.cn, cz277@tsinghua.edu.cn}
}

\maketitle

\begin{abstract}
Speech quality assessment typically requires evaluating audio from multiple aspects, such as mean opinion score (MOS) and speaker similarity (SIM) \textit{etc.}, which can be challenging to cover using one small model designed for a single task. In this paper, we propose leveraging recently introduced auditory large language models (LLMs) for automatic speech quality assessment. By employing task-specific prompts, auditory LLMs are finetuned to predict MOS, SIM and A/B testing results, which are commonly used for evaluating text-to-speech systems. Additionally, the finetuned auditory LLM is able to generate natural language descriptions assessing aspects like noisiness, distortion, discontinuity, and overall quality, providing more interpretable outputs.
Extensive experiments have been performed on the NISQA, BVCC, SOMOS and VoxSim speech quality datasets, using open-source auditory LLMs such as SALMONN, Qwen-Audio, and Qwen2-Audio. For the natural language descriptions task, a commercial model Google Gemini 1.5 Pro is also evaluated. The results demonstrate that auditory LLMs achieve competitive performance compared to state-of-the-art task-specific small models in predicting MOS and SIM, while also delivering promising results in A/B testing and natural language descriptions. Our data processing scripts and finetuned model checkpoints can be found at \url{https://github.com/bytedance/SALMONN}.
\end{abstract}
\begin{IEEEkeywords}
Speech quality assessment, auditory LLM, multimodal LLM, mean opinion score, speaker similarity
\end{IEEEkeywords}

\section{Introduction}
With the rapid development of generative models, the task of automatic speech quality assessment has become increasingly more important, for purposes from training data filtering to evaluating text-to-speech (TTS) models \cite{automos,mosnet,utmos,ssl-mos,voicemos2024}. However, considering the variety of aspects humans take into account in listening tests, it is challenging for small models trained for the end-to-end prediction of mean opinion score (MOS) \cite{mosnet,utmos} and speaker similarity score (SIM) \cite{voxsim} to achieve as comprehensive and interpretable speech quality assessment as humans. Recently, auditory large language models (LLMs) \cite{audiopalm,salmonn,qwenaudio,wavllm,qwen2audio,gemini} have been developed and demonstrated remarkable performance across a variety of speech perception and understanding tasks \cite{yassir,wujian,wenyi,bat}, which are thus possible to achieve evidence-based and multi-perspective speech quality assessment using natural language. 

In this paper, we propose to enhance existing auditory LLMs with the ability to evaluate speech quality from two key perspectives: MOS and SIM. Under an LLM multitask setting empowered by task-specific prompts, the auditory LLMs are evaluated using four speech quality evaluation tasks. These tasks include the predictions of real-valued quantitative metrics like MOS and SIM and an A/B testing to choose the better one from two speech samples, which are commonly used in TTS performance evaluation. Furthermore, we explore natural-language-based speech quality evaluation tasks, such as speech descriptions from the perspectives of noisiness, distortion, discontinuity, and overall quality. These tasks leverage the key strengths of the LLM backbone and are unique abilities of auditory LLMs. Experimental results on the widely used NISQA, BVCC, SOMOS and VoxSim show that finetuned auditory LLMs can serve as a versatile evaluation model for speech quality, outperforming a strong self-supervised learning (SSL) baseline and being competitive to state-of-the-art results achieved using separate small models.


\section{Related Work}

\subsection{Auditory LLM for Diverse Comprehension Tasks}
LLMs \cite{llama,gpt4,gemini} have shown exceptional abilities in text processing. By integrating an audio encoder with a text-based LLM, auditory LLMs establish a versatile framework for tackling audio-to-text tasks \cite{audiopalm,salmonn,qwenaudio,wavllm,tang2024extending}. Models like SALMONN \cite{salmonn}, Qwen-Audio \cite{qwenaudio}, and Google Gemini \cite{gemini} aim to unify audio perception and understanding tasks using a single model.
Other works focus on enhancing auditory LLMs for specific tasks, such as automatic speech recognition \cite{yassir,wenyi,embarrass}, speech translation \cite{wujian,salm}, and spoken language understanding \cite{slu}. In the meantime, efforts have been made to expand the range of tasks auditory LLMs can handle, such as spatial audio processing \cite{bat,spatial} and audio entailment \cite{audioentail}. 
To evaluate the strengths and limitations of auditory LLMs, evaluation benchmarks like DynamicSUPERB \cite{superb}, AIR-Bench \cite{airbench}, and AudioBench \cite{audiobench} have been developed, which are not only to assess the existing abilities of auditory LLMs, but also to suggest the necessary abilities to build.
Along these lines, this paper contributes to this growing field by pushing the boundaries of auditory LLMs in speech quality assessment. 


\begin{figure*}[t]

\centering
\begin{minipage}[b]{0.45\linewidth}
  \centering  \hspace{-0.5cm}\centerline{\includegraphics[width=1.1\linewidth]{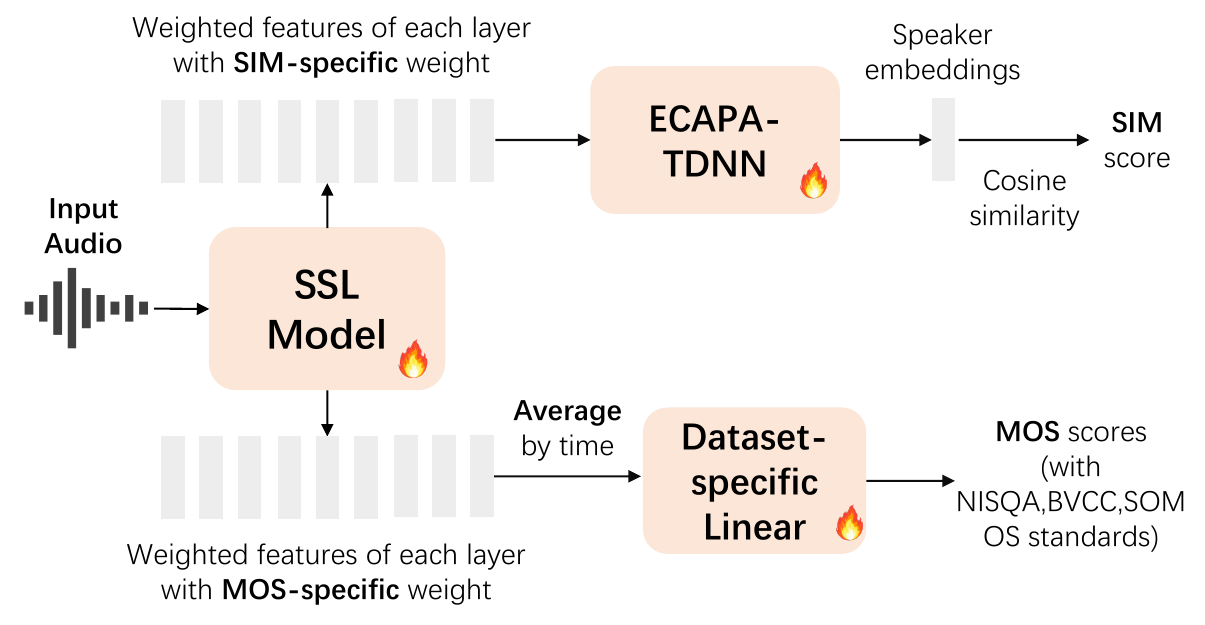}}
  \centerline{\scriptsize{(a) One SSL model for MOS and SIM prediction.}}\medskip
\end{minipage}
\hspace{0.8cm}
\begin{minipage}[b]{0.45\linewidth}
  \centering  
  \centerline{\includegraphics[width=1.2\linewidth]{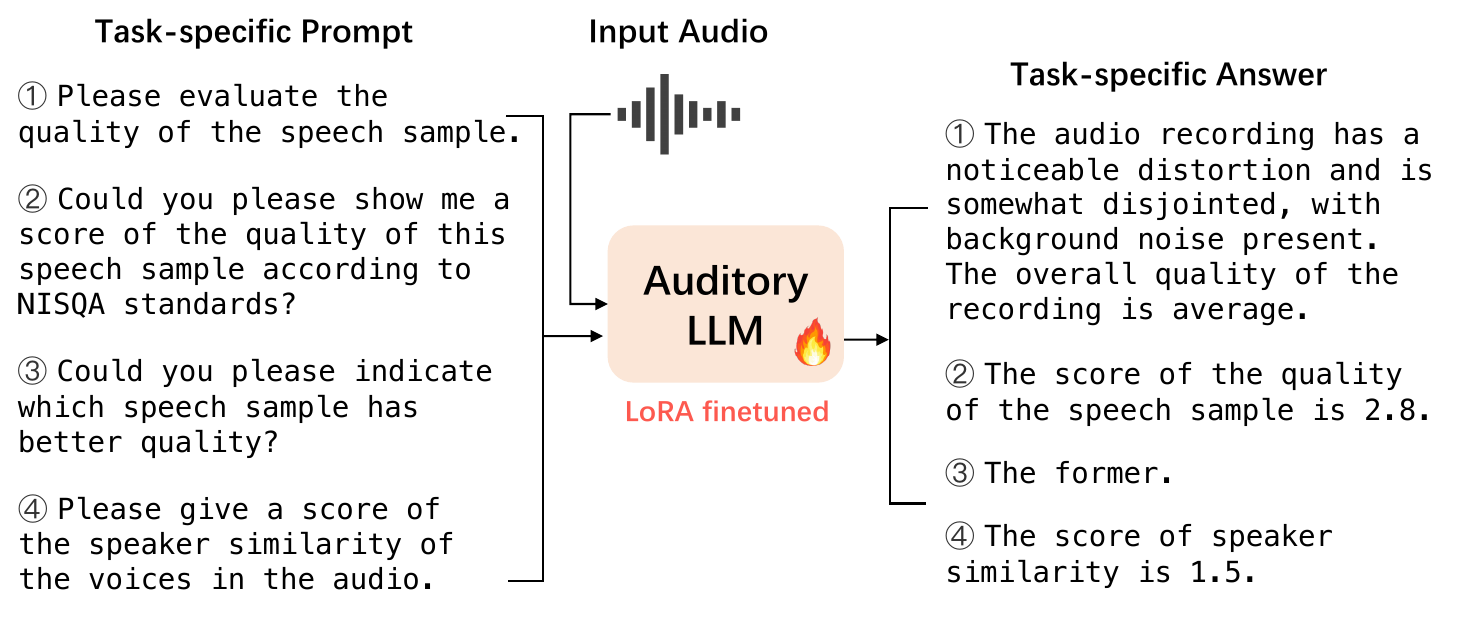}}
  \vspace{0.44cm}
  \centerline{\scriptsize{(b) Speech quality assessment auditory LLM.}}\medskip
\end{minipage}
\vspace{-0.20cm}
\caption{Sketchmaps of the speech quality assessment models. (a) The speech SSL model baseline. The SSL model and downstream models are fully finetuned using the speech quality assessment dataset, involving MOS and SIM prediction. (b) Speech quality assessment auditory LLM. The auditory LLM is finetuned with LoRA on the speech quality assessment dataset. Task-specific prompts are used to enable auditory LLMs to perform different speech assessment tasks, including MOS/SIM prediction, speech quality A/B testing and natural language descriptions.}
\label{fig1}
\end{figure*}
\vspace{-0.05cm}

\subsection{Automatic Speech Quality Evaluation}

Recent advancements in automatic speech quality evaluation have primarily focused on predicting MOS, which is often obtained by averaging discrete opinion scores assigned by dozens of independent human listeners. AutoMOS \cite{automos} employs a deep recurrent neural network to predict MOS, achieving correlations closely aligned with human evaluators. MOSNet \cite{mosnet} explores different model architectures, identifying CNN+BLSTM as the top performer. SSL-MOS \cite{ssl-mos} further enhances accuracy by utilizing pre-trained speech SSL models. UTMOS \cite{utmos}, based on ensemble learning of strong and weak learners, achieved the highest scores on several metrics in the main track of the VoiceMOS Challenge 2022 \cite{voicemos}. Moreover, the recently developed speech language models have been applied to provide alternative perspectives on synthesized speech evaluation, showing promising correlations with subjective metrics \cite{speechlmscore,speechbertscore,pam,llm2}. 

SIM, a key metric in voice conversion and zero-shot text-to-speech TTS evaluation, is often predicted using speaker verification models. The cosine distance between speaker embeddings of the generated and reference speech is regarded as highly correlated with speaker similarity. Small models (models with fewer than 1 billion parameters \cite{scalinglaw}) like ECAPA-TDNN \cite{ecapa} are typically used to extract these embeddings, and when combined with SSL features, models like WavLM-ECAPA \cite{wavlm} produce superior speaker embeddings. Modified MOSNet \cite{mosnet} has also shown the ability to predict SIM values with strong correlations to human ratings.

In this paper, we aim to develop a versatile speech quality evaluation model based on auditory LLMs, capable of predicting both MOS and SIM and performing A/B testing, while also assessing the quality of a speech sample by grounding the natural language descriptions on auditory evidence from multiple perspectives.


\section{Methodology}

\subsection{Auditory LLMs for Audio Perception and Understanding}
Auditory LLMs typically consist of three key components: an audio encoder, a connection module, and an LLM. The audio encoder extracts features from the input audio, which are then processed by the connection module to align the audio encoder output space with the LLM textual input space. The aligned features are then concatenated with the embeddings of the text prompt to form the input for the auditory LLM, which generates responses based on the input audio and text prompt. A multi-stage training approach is used, starting with a pre-training phase to align auditory and textual information, followed by an instruction-tuning phase to improve the model’s ability to handle more complex comprehension tasks. 
Regarding the trainable parameters, in SALMONN, both the audio encoder and the LLM are frozen, with only the connection module and low-rank adaptation (LoRA) \cite{lora} in the LLM being trained. LoRA is a parameter-efficient finetuning method for LLM. In contrast, Qwen-Audio also freezes the LLM but does not use LoRA; instead, the audio encoder is finetuned to derive the audio features in the textual input space of the LLM backbone. In this paper, we experiment with the open-source auditory LLMs, including SALMONN \cite{salmonn}, Qwen-Audio \cite{qwenaudio} and Qwen2-Audio \cite{qwen2audio}.

SALMONN \cite{salmonn}  This design enables SALMONN to effectively interpret a wide range of  general audio inputs. The outputs from both encoders are concatenated and fed into a connection module, a window-level Q-Former, which shares the same architecture as the vanilla image Q-Former \cite{blip2} but operates at the window level. Instead of processing the entire audio at once, the window-level Q-Former generates one textual token for approximately every 0.33 seconds, resulting in 88 tokens for a 30-second audio. The LLM backbone is the Vicuna LLM \cite{vicuna} adapted by LoRA, which generates responses with audio textual tokens and text prompt tokens presenting.

Qwen-Audio \cite{qwenaudio} has a simplified model structure, which contains an audio encoder and an LLM backbone. Only a pooling layer with a stride of two is incorporated to reduce the length of the audio representation. The audio encoder is initialised by the Whisper-large-v2 \cite{whisper} encoder, which is fully finetuned during multi-task training to improve the extraction of non-speech audio information. The LLM is a frozen Qwen-7B \cite{qwen}. Qwen2-Audio \cite{qwen2audio} follows the same model architecture of Qwen-Audio but improves upon both the audio encoder and the LLM. Reinforcement learning is also introduced to train the model to align with human preferences.


\begin{table*}[t]
\caption{MOS prediction results on test split of NISQA, BVCC and SOMOS datasets. For NISQA, the average results of the three public available test sets (NISQA\_TEST\_LIVETALK, NISQA\_TEST\_FOR and NISQA\_TEST\_P501) are reported. No system-level results for NISQA because there is no system label in the NISQA dataset.}
\setlength{\tabcolsep}{5pt}
\centering
\begin{tabular}{c|ccc|cccccc|cccccc}
\toprule
\multirow{3}{*}{\makecell{Model}} & \multicolumn{3}{c|}{NISQA} & \multicolumn{6}{c|}{BVCC} & \multicolumn{6}{c}{SOMOS} \\
 & \multicolumn{3}{c|}{utterance-level} & \multicolumn{3}{c}{utterance-level} & \multicolumn{3}{c|}{system-level} & \multicolumn{3}{c}{utterance-level} & \multicolumn{3}{c}{system-level}  \\
 & LCC & SRCC & MSE & LCC & SRCC & MSE & LCC & SRCC & MSE & LCC & SRCC & MSE & LCC & SRCC & MSE \\
\midrule
\textcolor{gray}{Single-task SOTA} & \textcolor{gray}{0.894} & \textcolor{gray}{0.887} & \textcolor{gray}{0.286} & \textcolor{gray}{0.899} & \textcolor{gray}{0.896} & \textcolor{gray}{0.165} & \textcolor{gray}{0.939} & \textcolor{gray}{0.936} & \textcolor{gray}{0.090} & \textcolor{gray}{0.687} & \textcolor{gray}{0.681} & \textcolor{gray}{0.203} & \textcolor{gray}{0.911} & \textcolor{gray}{0.917} & \textcolor{gray}{0.052} \\
WavLM-Base+ & 0.739 & 0.724 & 0.536 & 0.769 & 0.765 & 0.350 & 0.805
& 0.799 & 0.251 & 0.349 & 0.318 & 0.288 & 0.544 & 0.539 & 0.081 \\
WavLM-Large & 0.850 & 0.845 & 0.467 & 0.813 & 0.805 & 0.377 & \textbf{0.884} & 0.879 & 0.189 & 0.623 & 0.613 & 0.257 & 0.880 & 0.880 & 0.044\\
\midrule
SALMONN (vic1.5) & {\textbf{0.861}} & {\textbf{0.859}} & {\textbf{0.347}} & {\textbf{0.826}} & {\textbf{0.833}} & {\textbf{0.282}} & {\textbf{0.884}} & {\textbf{0.884}} & {\textbf{0.152}} & {\textbf{0.644}} & {\textbf{0.636}} & {\textbf{0.196}} & {0.894} & {0.891} & {0.034} \\

SALMONN (vic1.0) & {0.829} & {0.831} & {0.453} & {0.819} & {0.827} & {0.296} & {0.860} & {0.864} & {0.181} & {0.631} & {0.626} & {0.200} & {\textbf{0.902}} & {\textbf{0.904}} & {\textbf{0.030}} \\
Qwen-Audio & 0.771 & 0.784 & 0.664 & 0.680 & 0.685 & 0.451 & 0.790 & 0.803 & 0.375 & 0.569 & 0.553 & 0.225 & 0.834 & 0.858 & 0.036\\
Qwen2-Audio & 0.768 & 0.780 & 0.643 & 0.681 & 0.678 & 0.493 & 0.800 & 0.797 & 0.247 & 0.583 & 0.572 & 0.216 & 0.850 & 0.873 & 0.040\\
\bottomrule
\end{tabular}
\label{t1}
\end{table*}

\subsection{Speech Quality Evaluation with Auditory LLM}
The exceptional text processing abilities enable auditory LLMs to unify different speech quality assessment tasks, simply by generating real-valued numbers for MOS and SIM predictions, and words as text strings for A/B testings and natural language descriptions for speech quality. 
However, since current open-source auditory LLMs struggle to follow instructions accurately when exposed to untrained tasks \cite{salmonn}, the models are finetuned to increase their success rates in assessing speech quality. In this work, we focus on four different tasks of speech quality evaluation, namely MOS/SIM prediction, speech quality A/B testing and natural language descriptions. The task-specific prompts are shown in Fig \ref{fig1} (b). For each task, we employ three prompts with similar meanings during finetuning with one of them being randomly selected during the test.
The four tasks are formulated as follow:
\begin{itemize}
    \item \textbf{MOS prediction:} Auditory LLMs predict the MOS score on a scale of 1 to 5, for a given audio input. To improve performance across different datasets, dataset-specific prompts are designed. These prompts distinguish between datasets by incorporating the phrase ``\textit{according to {Dataset} standards}'', allowing the model to adjust its predictions based on the particular characteristics of each dataset.
    \item \textbf{Speech quality A/B testing:} The input has two speech samples with identical text from diffrent systems, and the model determines which is better by responding such as ``\textit{The former}'' or ``\textit{The latter}''.
    \item \textbf{Natural language descriptions for speech quality:} Auditory LLMs provide a detailed evaluation of speech samples, assessing aspects such as noisiness, distortion, discontinuity, and overall quality. An example response is: ``\textit{The audio recording has a noticeable distortion and is somewhat disjointed, with background noise present. The overall quality of the recording is average.}''
    \item \textbf{SIM prediction:} Auditory LLMs predict the SIM score on a scale of 1 to 6, with precision to one decimal place, based on a comparison of two speech samples.
\end{itemize}

\section{Experiment Setup}

\subsection{Datasets}
\label{ssec:datasets}
The MOS datasets used in our paper are NISQA \cite{nisqa}, BVCC \cite{bvcc} and SOMOS \cite{somos}. For MOS prediction task, all three datasets are used, with the SOMOS-clean scores used for the SOMOS dataset. For the speech quality A/B testing, we selected pairs of speech samples with the same text from the SOMOS dataset to create an A/B testing dataset. The train, validation, and test splits contain 13,820, 2,260, and 2,212 speech samples, respectively, with no pair sharing the same MOS score. As for natural language descriptions for speech quality, evaluation descriptions are generated from the NISQA dataset based on scores for noisiness, distortion, discontinuity, and overall quality, following the rating descriptions in \cite{rating}. These descriptions are further diversified by rephrased using the LLama3-8B-Instruct LLM \cite{llama3}. The train, validation, and test set sizes for this task are 10,899, 2,635, and 712 samples, respectively.

The SIM dataset used is VoxSim \cite{voxsim}, with average scores applied for evaluation. Since the original dataset lacks a valid split, we restructured the data by dividing the training set into a 9:1 ratio, yielding 22,389 samples for training and 2,532 for validation. For the speech quality A/B testing and SIM prediction tasks on SALMONN, a 2-second silence is introduced between concatenated speech samples, and any sample exceeding 14 seconds is truncated to that length.

\subsection{Models}

Four auditory LLMs are used in this paper, including SALMONN (vic1.5), SALMONN (vic1.0) \cite{salmonn}, Qwen-Audio \cite{qwenaudio} and Qwen2-Audio-7B \cite{qwen2audio}. The LLM backbones used by the SALMONN models are denoted in parentheses, where \textbf{vic1.5} denotes Vicuna-v1.5-7B and \textbf{vic1.0} denotes Vicuna-v1.0-13B. The connectors and LoRA on LLM backbones are finetuned, across all four tasks outlined in Sec.~\ref{ssec:datasets}, following the official scripts released\footnote{For SALMONN: https://github.com/bytedance/SALMONN; For Qwen-Audio: https://github.com/modelscope/ms-swift/issues/1653}. The LoRA rank is 8 and the scale is 32. The model is trained for 10 epochs, and the one that performs best on the validation set is selected for testing.

\subsection{Baselines}
Baselines include single-task SOTA and multipurpose baseline. The multipurpose baseline is built to compare more fairly in a multitask setting. The single-task SOTA baseline varies across datasets, which is NISQA \cite{nisqa} for NISQA dataset, UTMOS \cite{utmos} for BVCC dataset, modified SSL-MOS \cite{somosbest} for SOMOS dataset and WavLM-ECAPA \cite{wavlm} for VoxSim dataset. The multipurpose baseline is developed based on an SSL model for multipurpose speech quality assessment, focusing on MOS and SIM predictions. Since it is difficult for SSL models to generate word sequences, we do not enable the SSL model baseline to perform the task of natural language description for speech quality assessment. The model structure is shown in Fig \ref{fig1} (a). To combine the features from different layers of the SSL model for different downstream tasks, task-specific weights are learned for each layer. For SIM prediction, the downstream model is ECAPA-TDNN \cite{ecapa}, which is connected to the SSL model following the WavLM-ECAPA \cite{wavlm} configuration. The similarity of speaker embeddings generated by ECAPA-TDNN is linearly transformed to a range from 1 to 6 to match the speaker similarity scores in VoxSim \cite{voxsim}. For MOS prediction, the downstream model is a linear layer with time-averaged SSL features as input, similar to the setup in SSL-MOS \cite{ssl-mos}. Since using a single linear layer for MOS prediction across multiple datasets results in suboptimal performance, we implement dataset-specific linear layers to enhance accuracy. The model is jointly trained on MOS and SIM datasets for 20 epochs, with the best-performing model on the validation set selected as the baseline. Two baselines are built using WavLM-Base+ and WavLM-Large.

\subsection{Evaluation Criteria}

For MOS and SIM prediction, linear correlation coefficient (LCC), Spearman's rank correlation coefficient (SRCC) and mean square error (MSE) are calculated. For speech quality A/B testing, the accuracy is reported. For speech quality natural language assessment, the correlation score is evaluated by GPT-4o mini\footnote{The version is gpt-4o-mini-2024-07-18.}, given descriptive evaluation results and scores for noisiness, distortion, discontinuity, and overall quality.

\section{Results}

\subsection{MOS Prediction}

The MOS prediction results are presented in Table \ref{t1}. The SALMONN series models consistently match or surpass the performance of the WavLM-Large baseline across all three datasets, with SALMONN (vic1.5) standing out. We speculate SALMONN (vic1.5)'s superior performance to the improved quality of LLM, despite its smaller size. In contrast, the Qwen-Audio series models do not demonstrate competitive results. Interestingly, SALMONN (vic1.5) outperforms SALMONN (vic1.0) at the utterance level on SOMOS dataset, though this advantage does not extend to the system level, a pattern also observed in SSL models \cite{ssl-mos}. Since MOS prediction models are often used to evaluate the quality of a TTS system, system-level results may hold greater significance. Overall, these results suggest that finetuned auditory LLMs can effectively serve as MOS predictors.

The impact of dataset-specific prompts is also explored. Experiments on BVCC dataset using SALMONN (vic1.5), as shown in Table \ref{t2}, reveal that using a prompt tailored to the dataset yields the best system-level results, while averaging the scores from three different prompts performs best at the utterance level. This highlights both the effectiveness and generalization of dataset-specific prompts. As expected, the prompt corresponding to the dataset achieves the highest performance. For new, out-of-domain speech samples, averaging the results from multiple prompts may provide a more reliable predicted score, eliminating the need to create new prompts that might confuse the auditory LLM. Notably, using dataset-specific prompts with models not explicitly trained on them slightly hurts the performance, as observed in Table \ref{t2}. Given that the MOS score represents the average opinion of multiple individuals, an averaged predicted MOS score is likely to be more robust.

\begin{table}[h]
\caption{Ablation study of dataset-specific prompts for MOS prediction on BVCC dataset experimented with SALMONN (vic1.5).}
\setlength{\tabcolsep}{5pt}
\centering
\begin{tabular}{c|cccccc}
\toprule
\multirow{2}{*}{\makecell{Model}} & \multicolumn{3}{c}{utterance-level} & \multicolumn{3}{c}{system-level}\\
 & LCC & SRCC & MSE & LCC & SRCC & MSE \\
\midrule
\multicolumn{7}{c}{w/o dataset-specific prompt}  \\
w/o specific prompt & 0.827 & 0.826 & 0.324 & 0.874 & 0.874 & 0.228 \\
NISQA standards & 0.818 & 0.817 & 0.322 & 0.866 & 0.868 & 0.219 \\
BVCC standards & 0.824 & 0.824 & 0.349 & 0.873 & 0.873 & 0.249 \\
SOMOS standards & 0.823 & 0.822 & 0.336 & 0.868 & 0.869 & 0.239 \\
average 3 standards & 0.826 & 0.825 & 0.325 & 0.871 & 0.871 & 0.231 \\
\midrule
\multicolumn{7}{c}{with dataset-specific prompt}  \\
w/o specific prompt & 0.799 & 0.790 & 0.652 & 0.858 & 0.863 & 0.376 \\
NISQA standards & 0.813 & 0.819 & 0.533 & 0.861 & 0.861 & 0.311 \\
BVCC standards & 0.826 & 0.833 & \textbf{0.282} & \textbf{0.884} & \textbf{0.884} & \textbf{0.152} \\
SOMOS standards & 0.801 & 0.804 & 0.392 & 0.865 & 0.862 & 0.201 \\
average 3 standards & \textbf{0.829} & \textbf{0.835} & 0.340 & 0.879 & 0.878 & 0.174 \\
\bottomrule
\end{tabular}
\label{t2}
\end{table}

\subsection{SIM Prediction}

The SIM prediction results, listed in Table \ref{t3}, indicate that the SALMONN series models outperform the WavLM-Large baseline, with SALMONN (vic1.0) delivering the best performance. While the Qwen-Audio series models lag behind the WavLM-Base+ baseline. We don't report the results of Qwen-Audio, because the finetuned Qwen-Audio inaccurately predicts all SIM scores as 1. The results show that finetuned SALMONN can predict SIM well.

\begin{table}[h]
\caption{SIM prediction results on test split of VoxSim.}
\setlength{\tabcolsep}{5pt}
\centering
\begin{tabular}{c|ccc}
\toprule
Model & LCC & SRCC & MSE \\
\midrule
\textcolor{gray}{Single-task SOTA} & \textcolor{gray}{0.835} & \textcolor{gray}{0.836} & \textcolor{gray}{0.943} \\
WavLM-Base+ & 0.565 & 0.513 & 2.584 \\
WavLM-Large & 0.658 & 0.594 & 1.908 \\
\midrule
SALMONN (vic1.5) & 0.796 & 0.809 & 1.374 \\
SALMONN (vic1.0) & \textbf{0.816} & \textbf{0.824} & \textbf{1.199} \\
Qwen-Audio & - & - & -\\
Qwen2-Audio & 0.415 & 0.505 & 3.982 \\
\bottomrule
\end{tabular}
\label{t3}
\end{table}

\subsection{A/B testing}

The results of speech quality A/B testing are demonstrated on Table \ref{t4}.  In this task, the Qwen-Audio models outperform the SALMONN models, though the highest overall accuracy remains below 70\%. However, when focusing on cases where the MOS score difference between two speech samples exceeds 0.5, the accuracy of all auditory LLMs improves, with the best result reaching 80\%. This suggests that while auditory LLMs can compare speech quality to some extent, they struggle to differentiate between samples with similar scores.  Further improvements are necessary to achieve the accuracy required for practical applications.

\begin{table}[h]
\caption{Speech quality A/B testing results on test split of SOMOS. The numbers in the parentheses are the A/B testing accuracy values when the difference in MOS scores between the two speech samples exceeds 0.5.}
\setlength{\tabcolsep}{5pt}
\centering
\begin{tabular}{c|c}
\toprule
Model & Acc \\
\midrule
SALMONN (vic1.5) & 0.670 (0.761) \\
SALMONN (vic1.0) & 0.647 (0.739) \\
Qwen-Audio & 0.697 (\textbf{0.805}) \\
Qwen2-Audio & \textbf{0.698} (0.803) \\
\bottomrule
\end{tabular}
\label{t4}
\end{table}

\subsection{Natural Language Descriptions for Speech Quality}

The results of natural language descriptions are presented in Table \ref{t5}. The powerful commercial multimodal LLM Gemini 1.5 Pro \cite{gemini} is also tested\footnote{The version is Gemini-1.5-pro-preview on 2024.09.10.}. Pretrained open-source auditory LLMs often produce results that lack diversity or relevance to the task, leading to weaker performance. In contrast, finetuned auditory LLMs show a stronger understanding of prompts and generate more appropriate descriptive assessments, reflected in their improved correlation scores. However, the generated text remains somewhat monotonous. We believe that developing a dedicated dataset for natural-language-description-based speech quality assessment would allow auditory LLMs to evaluate speech quality more precisely. Notably, Gemini performs best among the pretrained LLMs, showcasing its strong capabilities. We anticipate even better results from Gemini with more carefully crafted prompts.

\begin{table}[h]
\caption{Results of natural language descriptions for speech quality on the test split of NISQA. The correlation score is calculated by GPT-4o-mini.}
\setlength{\tabcolsep}{5pt}
\centering
\begin{tabular}{c|cc}
\toprule
\multirow{2}{*}{\makecell{Model}} & \multicolumn{2}{c}{Correlation score} \\
 & pretrained & finetuned \\
\midrule
SALMONN (vic1.5) & 0.10 & \textbf{0.64} \\
SALMONN (vic1.0) & 0.11 & 0.60 \\
Qwen-Audio & 0.22 & \textbf{0.64} \\
Qwen2-Audio & 0.20 & 0.55 \\
Gemini 1.5 Pro & 0.43 & -- \\
\bottomrule
\end{tabular}
\label{t5}
\end{table}

\section{Conclusion}
In this paper, we propose to enhance pre-trained auditory LLMs to achieve general speech quality assessment tasks including MOS and SIM predictions, selecting the better of two speech samples through A/B testing, and evidence-based multi-aspect natural language descriptions for speech quality. Experimental results show that finetuned auditory LLMs can effectively serve as general-purpose speech quality evaluators and achieve competitive results compared to state-of-the-art task-specific models. 

\clearpage
\bibliographystyle{IEEEtran}
\bibliography{refs}

\end{document}